\documentclass[longauth]{aa}  
\usepackage{graphicx}
\usepackage{txfonts}
\usepackage{rotating} 
\usepackage{lscape}
\usepackage{xcolor}
\usepackage[colorlinks=true,linkcolor=blue,citecolor=blue,urlcolor=blue]{hyperref}

\usepackage{multirow}
\usepackage{adjustbox}
\usepackage{pifont}  
\usepackage{subcaption}

\newcommand{\tick}{\ding{51}} 

\newcommand{\kms}{\,km\,s$^{-1}$}
\newcommand{\msun}{\,M$_\odot$}
\newcommand{\NOEMA}{NOEMA$^{\mathrm{3D}}\,$}
\newcommand{\Reff}{R$_\mathrm{e}$}

\defcitealias{Chen2026}{Paper~2}

\begin{document}

   \title{NOEMA$^{\mathrm{3D}}$: Resolving radial gas flows in disk galaxies at $z\sim1.1-1.6$ with high-resolution CO observations}

   %\subtitle{I. Overviewing the $\kappa$-mechanism}
\titlerunning{NOEMA$^{\mathrm{3D}}$: Resolving radial gas flows at $z\sim1-1.5$}

  \author{  Jean-Baptiste Jolly \inst{1} \thanks{ \email{jbjolly@mpe.mpg.de}} \and
              Linda J. Tacconi \inst{1} \and
            Reinhard Genzel \inst{1,2} \and
            Roberto Neri \inst{3} \and
             Karl Schuster \inst{3} \and 
             Jianhang Chen \inst{1} \and
             Natascha M. F{\"o}rster Schreiber \inst{1} \and
             Stavros Pastras \inst{1,4} \and
             Letizia Scaloni \inst{5,6} \and
             Giulia Tozzi \inst{1} \and
              Capucine Barféty \inst{1} \and
              Alberto Bolatto \inst{7,8} \and
           Andreas Burkert\inst{1,9} \and
            Françoise Combes\inst{10} \and
             Pierre Cox\inst{11} \and
             Ric Davies \inst{1}\and
             Frank Eisenhauer \inst{1,12} \and
             Juan Manuel Espejo Salcedo \inst{1}\and
             Rodrigo Herrera-Camus\inst{13,14} \and
             Santi García-Burillo\inst{15} \and
             Tadayuki Kodama\inst{16} \and 
             Lilian Lee \inst{1}\and
             Minju M. Lee \inst{17,18} \and
             Daizhong Liu \inst{19} \and
             Dieter Lutz \inst{1}\and
             Giovanni Mazzolari \inst{1} \and
             Thorsten Naab \inst{4} \and
             Amit Nestor Shachar \inst{20} \and
             Claudia Pulsoni \inst{1}\and
             Alvio Renzini \inst{21} \and
             Monica Rubio \inst{22} \and
             Taro T. Shimizu \inst{1} \and
             Amiel Sternberg \inst{1,20,23}\and
             Eckhard Sturm \inst{1} \and
             Hannah  Übler \inst{1} \and
             Antonio Usero\inst{12} \and
             Stijn Wuyts\inst{24}        
         }

   \institute{Max-Planck-Institut für extraterrestrische Physik, 85748 Garching, Germany \\
              \email{jbjolly@mpe.mpg.de}
              \and Departments of Physics and Astronomy, University of California,
Berkeley, CA 94720, USA
         \and
             Institut de Radioastronomie Millimétrique (IRAM), 300 Rue de la Piscine, 38400 Saint-Martin-d’Hères, France
             \and 
             Max-Planck-Institut für Astrophysik (MPA), Karl-Schwarzschild-Str. 1, D-85748 Garching, Germany 
             \and 
             Department of Physics and Astronomy “Augusto Righi”, University of Bologna, Via Piero Gobetti 93/2, 40129 Bologna, Italy 
             \and
             INAF – Astrophysics and Space Science Observatory of Bologna, Via Piero Gobetti 93/3, 40129 Bologna, Italy
             \and
             Department of Astronomy, University of Maryland, College Park, MD 20742, USA
             \and
            Joint Space-Science Institute, University of Maryland, College Park, MD 20742, USA
            \and
             University Observatory, Ludwig Maximilians University, Scheinerstr. 1, 81679 Munich, Germany
             \and
             Observatoire de Paris, LUX, Collège de France, CNRS, PSL University, Sorbonne University, 75014 Paris, France
             \and
             Sorbonne Université, UPMC Paris 6 and CNRS, UMR 7095, Institut d’Astrophysique de Paris, 98b bd. Arago, 75014 Paris, France
             \and
             Technical University of Munich, TUM School of Natural Sciences, Physics Department, 85747 Garching, Germany
             \and
             Departamento de Astronomía, Universidad de Concepción, Barrio Universitario, Concepción, Chile
             \and
             Millenium Nucleus for Galaxies (MINGAL), Concepción, Chile 
             \and
             Observatorio Astronómico Nacional (OAN-IGN)-Observatorio de Madrid, Alfonso XII, 3, 28014 Madrid, Spain
             \and
             Astronomical Institute, Tohoku University, 6-3 Aramaki, Aoba-ku, Sendai 980-8578, Japan
            \and
                Cosmic Dawn Center (DAWN), Copenhagen, Denmark
            \and
                DTU-Space, Technical University of Denmark, Elektrovej 327, DK2800 Kgs. Lyngby, Denmark
            \and Purple Mountain Observatory, Chinese Academy of Sciences, 10 Yuanhua Road, Nanjing 210023, China
            \and
                School of Physics and Astronomy, Tel Aviv University, Tel Aviv 69978, Israel 
                \and
                Osservatorio Astronomico di Padova, Vicolo dell’Osservatorio 5, Padova I-35122, Italy
            \and
            Departamento de Astronomía, Universidad de Chile, Casilla 36-D, Santiago, Chile
            \and
            Centre for Computational Astrophysics, Flatiron Institute, 162 5th Avenue, New York NY 10010, USA
            \and
            Department of Physics, University of Bath, Claverton Down, Bath, BA2 7AY, UK        
             }
    
    \date{Submitted to A\&A}         
   %\date{Submitted to A\&A}%; accepted March 16, 1997}

\abstract{We present NOEMA$^{\mathrm{3D}}$, a unique high-resolution study of purely molecular gas kinematics at $z\sim1.1$ to $1.6$, providing a dedicated view of cold gas dynamics at the late stages of the peak epoch of cosmic star formation. Using deep ($\gtrsim 20$~hr on source per target) IRAM--NOEMA CO observations of 10 massive ($10.45\leq \mathrm{log(M}_\star/\mathrm{M}_{\odot})\leq11.43$) ) main-sequence galaxies, complemented by high-resolution \textit{JWST} imaging, we resolve the molecular gas kinematics and morphology on kiloparsec scales. We find that all galaxies exhibit ordered rotation with moderate intrinsic turbulence (median $\sigma_0\sim 32\pm10$\kms, median $V_{\rm c}/\sigma_0 \sim8.6\pm2.9$), consistent with dynamically turbulent disks at late cosmic noon. After modeling the axisymmetric rotation with the forward-modeling code \texttt{DysmalPy}, we reveal spatially coherent velocity residuals in all but one more inclined system. The inferred in-plane non circular motions reach amplitudes of $\sim 50-100~{\rm km~s^{-1}}$, significantly larger than typically observed in local disk galaxies. Interpreting these non-circular motions as radial flows we find that the velocity residuals spatially coincide with non-axisymmetric structures --spiral arms and bars-- demonstrating a direct link between galaxy morphology and gas transport at $z \sim 1$--2. In spiral galaxies, the residual velocity patterns are typically dominated by inflows, while barred systems display an apparent inflow-outflow pattern, characteristic of in-plane bar-driven gas motions. We further find that the inferred molecular gas inflow rates are substantial, with a typical net inflow rate of the order of the star formation rate ($\overline{\dot{M}}\sim-50$ \msun/yr). This implies that spiral arms and bars at cosmic noon are highly efficient at funneling cold gas toward galaxy centers, perhaps driving the buildup of bulges and feeding central star forming regions and supermassive black holes.
}


   \maketitle
%
%-------------------------------------------------------------------

\nolinenumbers

\section{Introduction}

The epoch of cosmic noon ($z\sim1$--$3$) is a key phase in galaxy evolution. This epoch witnesses the peak of the star formation rate density (SFRd), but also black hole accretion rate, and --most important for this work-- of cosmic molecular gas mass density \citep{Madau2014,NFSWuytsReview,Tacconi2020,Walter2020}. After this peak, the SFRd drops rapidly, marking the transition to the present-day Universe, dominated by more quiescent, less active galaxies. The peak of molecular gas mass translates to a typical gas to total baryon fraction \citep[$f_{\mathrm{gas}}\sim30$--60\%, see][]{Tacconi2010,Tacconi2018,Mirka2020} much higher than that observed in local galaxies \citep[$f_{\mathrm{gas}}$ < 10\%, e.g.][]{Saintonge2017}.

With the advent of ground-based spectroscopic surveys with integral field units \citep[IFUs, e.g.][]{NFS2009,NFS2018,Epinat2009,Law2009,Wisnioski2015,Wisnioski2019,Harisson2017,Gillman2020}, detailed kinematic analyses of galaxies at $z=1$--$3$ became possible. By spatially and spectrally resolving the ionized gas line emission, such studies showed that the majority of galaxies at this epoch were rotating disks. However, the majority of galaxies at cosmic noon have been shown to exhibit high velocity dispersion
 \citep[$\sim 30$\,--$60$\kms, e.g.][]{Epinat2009,NFS2006,NFS2009,NFS2018,Genzel2006,Genzel2011,Kassin2007,Kassin2012,Wisnioski2015,Wisnioski2025,Uebler2019}, especially compared to their rotation velocities ($v/\sigma\sim1$--$10$). Such characteristics showcase a specific kinematic state, highlighting the major role of the gas and its turbulence in the disk support \citep{Burkert2009,Dekel2009}. Gravitational instabilities arising in these turbulent disks also lead to the frequent formation of clumps \citep[e.g.][]{Genzel2010,Genzel2011,Genzel2014,Romeo2010,Romeo2014,FoersterSchreiber2011a,FoersterSchreiber2011b,Guo2012,Guo2015,Guo2018,Newman2012b,Newman2012a,Wuyts2012}.

The prevalence of disk galaxies at cosmic noon, together with the tight scaling relations of their properties, indicates that galaxy evolution at cosmic noon is governed mainly by internal secular processes and relatively smooth gas accretion, while major mergers play a secondary role. While major mergers were proposed, for a time, as the main drivers of mass growth at intermediate redshifts, they are now known to represent only a small fraction ($\sim 2-10$\,\%) of the observed galaxies at cosmic-noon \citep{Rodighiero2011,Ownsworth2014,Lofthouse2017,Mantha2017,Cibinel2019}. Consequently, the remaining channels of galaxy growth must be softer, less destructive, like minor mergers or continuous gas accretion from the intergalactic medium \citep[IGM,][]{Keres2005,Dekel2009,Genel2009,Genel2012,Genel2014,Martin2017}. 

\vspace{3mm}
{\noindent \textbf{Radial flows as a key mechanism in galaxy growth}}
\vspace{3mm}

In this context, a central prediction of modern galaxy formation models is that the sustained growth of galaxies is largely driven by the continuous accretion of gas from the circumgalactic and intergalactic media (CGM/IGM; e.g. \citealt{Keres2005,Dekel2009,Genzel2010,Genzel2015,Tumlinson2017}). This framework implies that the high gas fractions and elevated star formation rates observed at $z\sim1$--$3$ must be sustained by efficient gas inflows from large scales. A key open question is therefore how this accreted material is transported from the outer regions of galaxies down to the central, star-forming disk, and whether this process can account for the elevated gas fractions and star formation rates observed at cosmic noon.

Recent observational studies have begun to directly address this question, enabled by the high spectral and spatial resolution of modern ground-based millimeter interferometers, which can resolve gas kinematics on sub-kpc scales with velocity precision of a few tens of \kms. In particular, based mostly on H$\alpha$ data, \citet[][]{Genzel2023} reported the detection of significant non-circular motions in a sample of 9 star-forming galaxies at $z\sim1$--$2$, with velocities of several tens of \kms, interpreted as signatures of inward radial gas flows. Their detailed case studies indicate that substantial radial gas transport can occur within galactic disks, suggesting that angular momentum redistribution and gas inflow play an important role in shaping the internal baryon cycle, although the physical drivers of these flows are still being actively investigated. These results provide some of the first direct observational evidence linking large-scale gas accretion to the internal redistribution of gas within galaxies at cosmic noon, and highlight radial flows as a key mechanism connecting cosmological inflows to the regulation of star formation (see also Pulsoni et al., in prep.).

\vspace{3mm}
{\noindent \textbf{A novel view of galaxy morphology with \textit{JWST}}}
\vspace{3mm}

Since the end of 2022 and the progressive public release of data from the James Webb Space Telescope (\textit{JWST}), our view of galaxy morphology at cosmic noon and beyond has undergone a qualitative shift. Unlike the Hubble Space Telescope (\textit{HST}), which primarily probed rest-frame ultraviolet to optical emission at these redshifts, \textit{JWST} observes galaxies in the rest-frame optical to near-infrared at comparable or better spatial resolution, increased sensitivity, and significantly reduced impact of dust extinction. As a result, structures that previously appeared clumpy, irregular, or featureless in \textit{HST} imaging are now revealed as well-ordered disks, often exhibiting clear spiral arms and bars \citep[see e.g.][]{Chen2022,Ferreira2023,Guo2023,Guo2025,LeConte2024,Kuhn2024,EspejoSalcedo2025,Geron2025,Kalita2026}. Many of these systems were previously interpreted as dynamically disturbed or merger-driven, highlighting how strongly morphological classifications at high redshift depended on wavelength and dust obscuration.

In agreement with earlier kinematic spectroscopic studies, morphological analyses exploiting \textit{JWST} data have shown that a large fraction of galaxies at $z=1$--$3$ are disk-dominated systems. For example, \citet{EspejoSalcedo2025}, based on the visual classification of $\sim1500$ massive star forming galaxies at $z\sim1$--$2$, find that more than $80$\% are disks, with $\sim50$\% showing spiral structure and $\gtrsim10$\% hosting bars (see also e.g. \citealt{Kartaltepe2023,Kuhn2024,LeConte2024,Guo2025}). %; Parlanti et al. 2026).

The prevalence of such dynamically organized structures at an epoch characterized by high gas fractions and strong turbulence raises important questions for galaxy evolution. In particular, if spiral arms and bars are already common at cosmic noon, what role do they play in regulating the transport and redistribution of gas within galaxies, and in linking large-scale gas accretion to star formation?

\vspace{3mm}
{\noindent \textbf{\NOEMA: A detailed look into radial flows}}
\vspace{3mm}

In this paper we present the complete set of deep, high-resolution NOEMA observations of the molecular gas kinematics traced by the CO in 10 massive galaxies at $z\sim1$--2, obtained as part of the \NOEMA large program. Building on recent studies that have established the presence of non-circular motions and radial gas flows in massive star-forming galaxies at cosmic noon \citep[][]{Genzel2023}, we carry out a homogeneous and detailed kinematic analysis of each galaxy in our sample, effectively doubling the number of cosmic-noon galaxies with such measurement. Our primary goal is to investigate the connection between these non-circular motions and the underlying morphological structures, and in particular to assess how spiral arms and bars may contribute to the redistribution of gas and the fueling of star formation.

In \S\ref{sec:data_and_sample} we introduce the sample, the data products, and their reduction. In \S\ref{sec:analysis} we present the kinematic modeling and the tools used to derive residual velocity fields and quantify radial flows. In \S\ref{sec:results} we present the results, including both the global kinematic properties of the sample and the detailed residual analysis. The implications of our findings, together with methodological limitations are discussed in \S\ref{sec:discussion}, while in \S\ref{sec:ccl} we summarize our conclusions. Throughout the paper, we adopt the Chabrier stellar initial mass function \citep{Chabrier2003} and a flat $\Lambda$CDM cosmology with H$_0$ = 70 km\,s$^{-1}$ and $\Omega_\text{m} = 0.3$.

Other papers in this series focus on the spatial distributions of gas and star formation (\citealt{Chen2026}, hereafter Paper 2), comparisons with high-resolution simulations (Pastras et al. 2025 and in prep.), and spatially resolved star formation relations and timescales (Tozzi et al., in prep.).

This work is based on observations carried out under project number L19MD with the IRAM NOEMA Interferometer. IRAM is supported by INSU/CNRS (France), MPG (Germany) and IGN (Spain).

%--------------------------------------------------------------------
\section{Data and Sample} \label{sec:data_and_sample}

\subsection{Sample selection and description} \label{sec:sample}

\NOEMA\, is a MPG-IRAM Observatory Program (MIOP) NOEMA large program aimed at resolving the molecular interstellar medium (ISM) distribution and kinematics in a well-characterized pool of star-forming galaxies (SFGs) at $z\sim1-2$. It extends the great success of the previous integrated survey on molecular gas properties at cosmic noon, PHIBSS \citep{Tacconi2013,Tacconi2018,Freundlich2019} and seeks to conduct a high-redshift equivalent of resolved CO surveys similar to those in the nearby Universe \citep[e.g.][]{Leroy2013}. 
 The sample was drawn from the 3D-HST catalog \citep{Brammer2012,Momcheva2016} in the northern Great Observatories Origins Deep Survey \citep[GOODS-North,][]{Giavalisco2004} and Extended Groth Strip \citep[EGS,][]{Davis2007} fields, where multi-wavelength high-resolution images and spectra are available from \textit{HST} \citep{Grogin2011,Koekemoer2011,Skelton2014}, and now also from \textit{JWST} \citep{CEERS,JADES}. The availability of these deep archival data provides constraints on the stellar component complementary to the cold gas from \NOEMA. Sources were selected based on the following criteria:
\begin{itemize}

\item The availability of accurate redshifts, typically from ground-based spectroscopic observations.

\item High stellar masses ($M_\star \sim 10^{10.5}$--$10^{11.5}$\,\msun, around the Schechter mass), to ensure that CO emission is not severely affected by photodissociation in low-metallicity environments.

\item Location on, or close to, the star-forming main sequence \citep[$\Delta(\mathrm{MS}) = \log(\mathrm{SFR}/\mathrm{SFR}_{\mathrm{MS}}) < \pm 0.6$ dex;][]{Speagle2014}, in order to focus on typical star-forming galaxies.

\item Large galaxy sizes (effective radii $R_\mathrm{e} > 3\,\mathrm{kpc}$), ensuring a sufficient number of resolution elements across the disk (typically 2--3 within one $R_\mathrm{e}$).

\end{itemize}

The expected molecular gas masses and corresponding CO fluxes were computed using the scaling relations presented in \citet{Tacconi2018}. Retaining only galaxies with expected on-source integration time less than 35 hours resulted in a final pool of 83 observable galaxies. This full sample of potential targets is designed to be representative of the general population of SFGs at $z\sim1-2$.

\begin{figure*}
    \centering
    \includegraphics[width=0.95\linewidth]{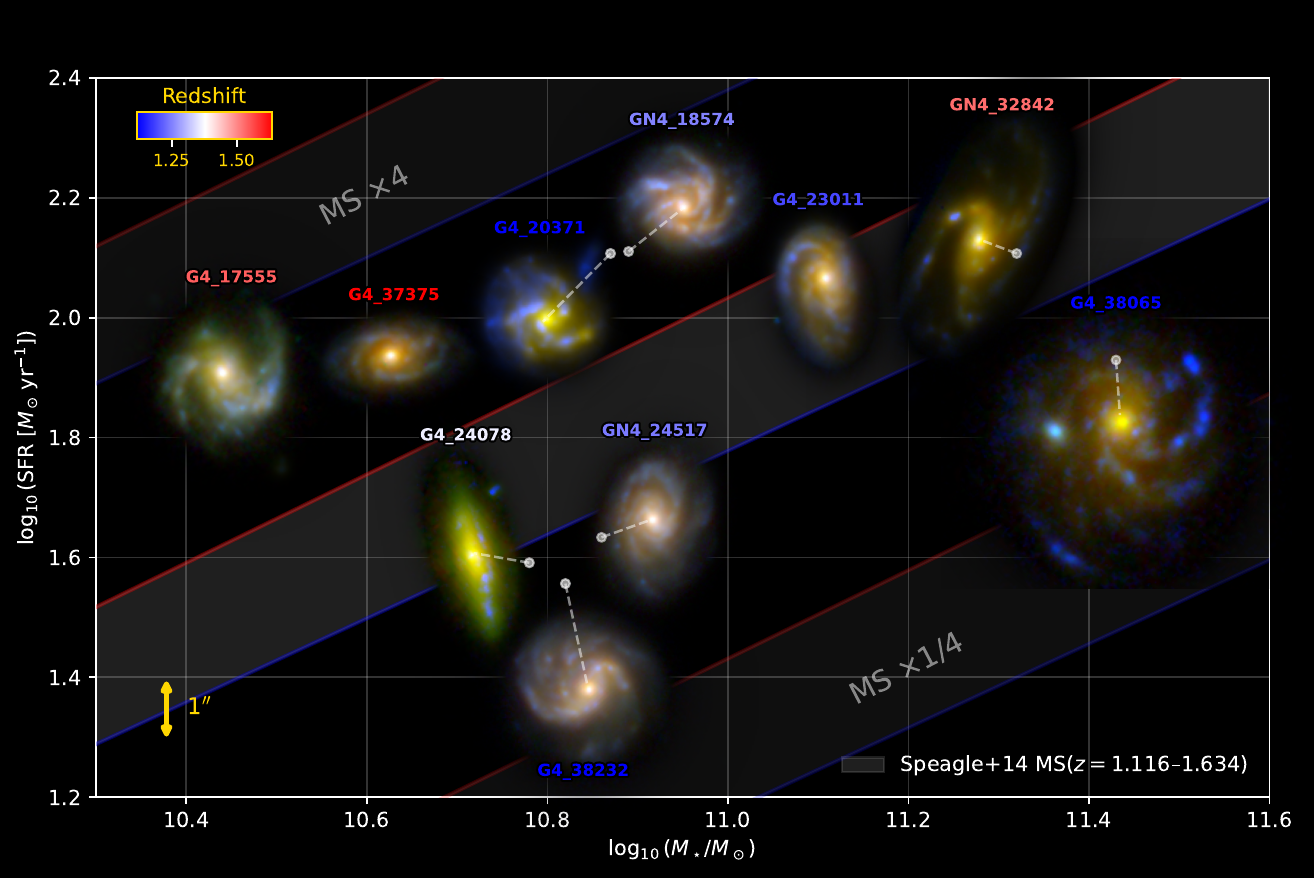}
    \caption{All galaxies in the \NOEMA sample: color composite images (typically from \textit{JWST}, NIRCam, see Table \ref{tab:photometry} for the list of filters used), placed on the SFR vs M$_\star$ plane (see Table \ref{tab:ppts} for the tabulated SFR and M$_\star$). To avoid overlap between galaxies, when needed, a gray dot, with a connecting dashed line, indicates the exact galaxy position. In the background, the main sequence of star forming galaxies \citep[MS, ][]{Speagle2014} is shown at the highest and lowest redshift of the sample (in red and blue respectively), and additional lines and corresponding shaded regions at $\mathrm{MS}\times4$ and $\mathrm{MS}\times1/4$ are also shown. The name of each galaxy is indicated above each, and color-coded for its redshift.}
    \label{fig:sample}
\end{figure*}

A total of $\sim30$ were observed first in compact NOEMA array configurations, for typical on-source integrations of a few hours, to establish firmly the source-integrated CO line fluxes and identify the best targets for high-resolution follow-up. From these, the 10 most promising sources were selected for follow up at high-resolution with deep integration. In Figure \ref{fig:sample} we show \textit{JWST}/\textit{HST} color composites of the sources, and their position on the SFR-M$_\star$ plane. Each target combines total NOEMA observations of typically more than 20 and up to 60 hours (see Table \ref{tab:obs}). Aside from one source these galaxies were selected for their moderate inclination and ordered morphological structures, showing spiral arms and, in four cases, bars. Four galaxies are observed in $^{12}$CO~$J=3-2$, hereafter CO(3--2) while the rest combine observations of $^{12}$CO~$J=4-3$, hereafter CO(4--3) and [C\,\textsc{i}](${}^3P_1-{}^3P_0$), hereafter[CI]. See \citetalias{Chen2026} for a detailed description of the different tracers. Redshifts vary from $z\sim1.12$ to $z\sim1.63$, the average stellar mass is 
$\overline{\mathrm{log(M_\star)}} \sim 10.92\pm0.27$, average effective radius $\overline{\mathrm{R_e}} \sim 4.53 \pm1.41$\,kpc and average star formation rate $\overline{\mathrm{SFR}} \sim 88\pm38$ \msun /yr. The properties of each galaxy are tabulated in Table \ref{tab:ppts}.

{
\setlength{\tabcolsep}{4pt}
\begin{table*}
\caption{\label{tab:ppts}General Galaxy Properties}
\centering
\begin{tabular}{lcccccccccccc}
\hline\hline
Name & $z$ & R$_\mathrm{e, star}$ & log(M$_\star$/M$_{\odot}$) & log(M$_\mathrm{gas}$/M$_{\odot}$) & f$_\mathrm{gas}$ & SFR & inc & PA & B/T & n$_{\mathrm{disk}}$ & Morphology \\
 & & (kpc) & (\msun) & (\msun) & & (\msun / yr) & (deg) & (deg) & &  &\\
\hline

G4\_38065 & 1.1151 & $8.3\pm1.3$ & $11.43\pm0.02$ & $10.82 \pm 0.03$ & 0.20 & 85 $\pm$ 4 & 18 & 12 & 0.09 & 1.1 & Spiral\\
G4\_38232 & 1.1159 & $4.4\pm0.3$ & $10.82\pm0.10$ & $10.65 \pm 0.08$ & 0.40 & 36 $\pm$ 15 & 20 & -66 & 0.16 & 1.0 & Barred\\
G4\_20371 & 1.1165 & $4.5\pm0.1$ & $10.87\pm0.05$ & $10.71 \pm 0.03$ & 0.41 & 128 $\pm$ 15 & 32 & -148 & 0.10 & 0.6 & Spiral\\
G4\_23011 & 1.1917 & $3.9\pm0.1$ & $11.10\pm0.04$ & $10.67 \pm 0.04$ & 0.28 & 121 $\pm$ 6 & 51 & 190 & 0.10 & 1.1 & Spiral \\
GN4\_24517 & 1.2411 & $3.7\pm0.1$ & $10.86\pm0.02$ & $10.47 \pm 0.06$ & 0.29 & 43 $\pm$ 2 & 41 & 154 & 0.10 & 1.0 & Barred\\
GN4\_18574 & 1.2463 & $3.7\pm0.1$ & $10.89\pm0.02$ & $10.86 \pm 0.02$ & 0.48 & 129 $\pm$ 6 & 12 & -40 & 0.06 & 0.6 & Sprial\\
G4\_24078 & 1.3595 & $4.7\pm0.2$ & $10.78\pm0.07$ & $10.50 \pm 0.09$ & 0.35 & 39 $\pm$ 12 & 76 & 14 & 0.07 & 1.4 & Edge-on\\
GN4\_32842 & 1.5233 & $5.1\pm0.1$ & $11.32\pm0.02$ & $10.80 \pm 0.06$ & 0.23 & 128 $\pm$ 6 & 49 & -20 & 0.09 & 1.5 & Barred\\
G4\_17555 & 1.5372 & $3.8\pm0.1$ & $10.45\pm0.12$ & $10.20 \pm 0.20$ & 0.36 & 83 $\pm$ 26 & 25 & 4 & 0.24 & 0.9 & Barred\\
G4\_37375 & 1.6335 & $3.2\pm0.1$ & $10.63\pm0.13$ & $10.20 \pm 0.30$ & 0.27 & 91 $\pm$ 30 & 50 & 100 & 0.23 & 0.9 & Spiral\\

\hline
\end{tabular}
\tablefoot{
Columns from left to right: galaxy identifier; redshift; effective radius of the galaxy  stellar light, derived from the curve-of-growth method, see \citetalias{Chen2026}; stellar mass from integrated SED fitting (see ; integrated molecular gas mass from NOEMA CO data (scaled following \citealt{Tacconi2020}, see \citetalias{Chen2026}); gas fraction ($M_{\mathrm{gas}}/(M_{\mathrm{*}}+M_{\mathrm{gas}})$); star formation rate from UV and IR luminosities, as resulting from integrated SED fitting; inclination (0\textdegree\ is face-on); kinematic position angle (positive east of north, pointing toward negative velocities); bulge-to-total mass ratio; Sérsic profile index of the disk; observed morphology. The \Reff, B/T and n$_{\mathrm{disk}}$ are extracted from curve of growth fitting and photometric fitting, see \citetalias{Chen2026}. 
}
\end{table*}
}

\subsection{Observations and data reduction} \label{sec:data}

The final datasets are a combination of observations in multiple configurations (typically A --the most extended NOEMA configuration, leading to the highest spatial resolution-- and C). Observations were carried out between 2019 and 2025. Sources at  $z<1.5$ were observed in NOEMA Band-3 (observing range from $\sim 196$ to $276$\,GHz, in $2\times7.744$\,GHz windows), targeting both the CO(4--3) and [CI] transitions, while sources at $z>1.5$ were observed in NOEMA Band-2 (observing range from $\sim 127$ to $\sim183$\,GHz, also in $2\times7.744$\,GHz windows), and targeting the CO(3--2) line. Each observation captured two polarizations. With the PolyFix correlator, this corresponds to an instantaneous bandwidth of $\sim16$\,GHz per polarization, enabling broad spectral coverage in each observation. All data were calibrated with \textsc{clic} and imaged with \textsc{mapping} from the GILDAS package \footnote{\url{https://www.iram.fr/IRAMFR/GILDAS/}} \citep{GILDAS}. \textsc{mapping} was systematically done with \textit{a)} a taper of 800m and a power of 2 (this configuration leading to the lowest resolution and increases sensitivity to large-scale emission), \textit{b)} natural weighting, and \textit{c)} uniform weighting with a robust factor of 5 (highest resolution, lowest sensitivity to large-scale emission). For sources with high S/N, maps with smaller robust factors (i.e. higher spatial resolution) were also created. Cleaning was applied only to channels with a S/N of at least 5, without the use of a clean mask.

\begin{sidewaystable*}

\caption{\label{tab:obs} Main properties of the data cubes.}
\centering   
 \begin{tabular}{lccccccccccccc}
 \hline \hline
 
Name&RA&Dec&z& 
Line&Freq&Config&$t_{\mathrm{int}}$&Weight&Beam size&Beam PA&Chan width&noise&$S_{\mathrm{CO}}$ \\ 
&\multicolumn{2}{c}{(J2000)}&&&GHz&&hours&&arcsec&degrees&MHz&mJy / beam&Jy km/s \\
\hline
\multirow{2}{*}{G4\_38065}&\multirow{2}{*}{14:20:05.500}&\multirow{2}{*}{53:01:15.60}&\multirow{2}{*}{1.1151}&\multirow{2}{*}{CO4-3}&\multirow{2}{*}{217.959}&\multirow{2}{*}{ACD}&\multirow{2}{*}{61.78}&NA&0.76x0.69&29&7&0.184&1.69\\
&&&&&&&&Ro5&0.51x0.42&37&7&0.241&0.82\\[3pt]

\multirow{2}{*}{G4\_38232}&\multirow{2}{*}{14:19:48.926}&\multirow{2}{*}{52:58:32.03}&\multirow{2}{*}{1.1159}&\multirow{2}{*}{CO4-3}&\multirow{2}{*}{217.866}&\multirow{2}{*}{AC}&\multirow{2}{*}{36.78}&NA&0.3x0.2&-166&20&0.121&0.81\\
&&&&&&&&Ro5&0.24x0.15&-162&20&0.13&0.51\\[3pt]

\multirow{2}{*}{G4\_20371}&\multirow{2}{*}{14:20:25.129}&\multirow{2}{*}{53:00:27.58}&\multirow{2}{*}{1.1165}&\multirow{2}{*}{CO4-3}&\multirow{2}{*}{217.835}&\multirow{2}{*}{AC}&\multirow{2}{*}{36.88}&NA&0.46x0.39&-159&20&0.115&1.37 \\
&&&&&&&&Ro5&0.24x0.16&-169&20&0.134&1.36\\[3pt]

\multirow{2}{*}{G4\_23011}&\multirow{2}{*}{14:19:11.206}&\multirow{2}{*}{52:48:00.35}&\multirow{2}{*}{1.1917}&\multirow{2}{*}{CO4-3}&\multirow{2}{*}{210.369}&\multirow{2}{*}{AC}&\multirow{2}{*}{28.07}&NA&0.45x0.34&29&20&0.138&1.64\\
&&&&&&&&Ro5&0.36x0.25&-158&20&0.141&1.60\\[3pt]

\multirow{2}{*}{GN4\_24517}&\multirow{2}{*}{12:36:21.34}&\multirow{2}{*}{62:15:46.00}&\multirow{2}{*}{1.2411}&\multirow{2}{*}{CO4-3}&\multirow{2}{*}{205.741}&\multirow{2}{*}{AC}&\multirow{2}{*}{23.18}&NA&0.76x0.66&-99&20&0.126&0.38\\
&&&&&&&&Ro5&0.63x0.57&-269&20&0.127&0.37\\[3pt]

\multirow{2}{*}{GN4\_18574}&\multirow{2}{*}{12:37:02.740}&\multirow{2}{*}{62:14:01.67}&\multirow{2}{*}{1.2463}&\multirow{2}{*}{CO4-3}&\multirow{2}{*}{205.244}&\multirow{2}{*}{AC}&\multirow{2}{*}{17.81}&NA&0.62x0.57&37&10&0.198&1.48\\
&&&&&&&&Ro5&0.53x0.48&42&10&0.2098&1.25\\[3pt]

\multirow{2}{*}{G4\_24078}&\multirow{2}{*}{14:19:15.996}&\multirow{2}{*}{52:49:10.54}&\multirow{2}{*}{1.3595}&\multirow{2}{*}{CO3-2}&\multirow{2}{*}{146.541}&\multirow{2}{*}{AC}&\multirow{2}{*}{22.22}&NA&0.91x0.78&-135&20&0.082&0.62\\
&&&&&&&&Ro5&0.77x0.63&-140&20&0.085&0.56\\[3pt]

\multirow{2}{*}{GN4\_32842}&\multirow{2}{*}{12:37:22.531}&\multirow{2}{*}{62:18:38.19}&\multirow{2}{*}{1.5233}&\multirow{2}{*}{CO3-2}&\multirow{2}{*}{137.057}&\multirow{2}{*}{AC}&\multirow{2}{*}{27.61}&NA&1.12x0.92&-105&20&0.067&1.04\\
&&&&&&&&Ro5&0.86x0.69&58&20&0.070&0.96\\[3pt]

\multirow{2}{*}{G4\_17555}&\multirow{2}{*}{14:19:19.684}&\multirow{2}{*}{52:48:14.49}&\multirow{2}{*}{1.5372}&\multirow{2}{*}{CO3-2}&\multirow{2}{*}{136.281}&\multirow{2}{*}{AC}&\multirow{2}{*}{26.70}&NA&1.18x0.96&33&10&0.102&0.28 \\
&&&&&&&&Ro5&0.71x0.47&-155&10&0.121&0.26\\[3pt]

\multirow{2}{*}{G4\_37375}&\multirow{2}{*}{14:19:06.731}&\multirow{2}{*}{52:50:39.68}&\multirow{2}{*}{1.6335}&\multirow{2}{*}{CO3-2}&\multirow{2}{*}{131.305}&\multirow{2}{*}{A}&\multirow{2}{*}{27.73}&NA&0.62x0.46&-33&20&0.069&0.36\\
&&&&&&&&Ro5&0.41x0.29&-14&20&0.077&0.37\\
\hline
\end{tabular}
\tablefoot{Columns from left to right: Name/identifier of the galaxy; Right Ascension and Declination \citep[from the 3D-HST catalog ][]{Brammer2012}; redshift (from the CO line); main targeted line; central frequency; observing configuration of the array: A (D) corresponds to the most (least) extended configuration, and hence the highest (lowest) resolution; total integration time on source (12 antennas equivalent); visibility weights, NA is natural, Ro5 is robust with a robust parameter of 5 in the GILDAS software definition\footnote{See e.g. \url{https://www.iram.fr/IRAMFR/GILDAS/doc/pdf/map.pdf}}; beam size (major axis $\times$ minor axis); position angle of the beam, positive north to east; adopted spectral channel width; noise per channel; CO line flux.
}
\end{sidewaystable*}

\subsection{Ancillary photometric data and SED fitting} \label{sec:SED}

We collected publicly available \textit{HST} and \textit{JWST} imaging from various surveys, including the Cosmic Assembly Near-infrared Deep Extragalactic Legacy Survey (CANDELS; \citealt{Grogin:2011,Koekemoer:2011}), the \textit{JWST} Advanced Deep Extragalactic Survey (JADES; \citealt{Eisenstein:2023}), the Cosmic Evolution Early Release Science Survey (CEERS; \citealt{Finkelstein:2023}), and the First Reionization Epoch Spectroscopically Complete Observations (FRESCO; \citealt{Oesch:2023}). When available, we retrieved high-quality reduced \textit{JWST} data from the public DAWN \textit{JWST} Archive (DJA). Otherwise, we retrieved the raw data from the MAST archive and reduced them using a pipeline built around the \texttt{CrabToolkit} \footnote{\url{https://github.com/1054/Crab.Toolkit.JWST}} following the standard reduction steps with additionally improved "snowball", "claw" and "wisp" using publicly available templates \footnote{\url{https://jwst-docs.stsci.edu/known-issues-with-jwst-data/nircam-known-issues/nircam-scattered-light-artifacts}}, application of astrometric corrections based on the CEERS photometric catalog \citep{Cox_2025}, as well as custom masking.

Multi-band \textit{JWST}/NIRCam imaging is available for all sources but one (G4\_38065), and six targets also have MIRI imaging (see Table \ref{tab:photometry}). The \textit{HST} data consist of ACS and WFC3/IR images, covering F435W to F160W wavebands ($\sim$\,0.4\,--\,1.6 $\mu$m). For galaxies with \textit{JWST}/NIRCam and \textit{HST} filters covering a similar wavelength range (e.g. \textit{JWST}/NIRCam F150W and \textit{HST}/WFC3 F160W), we only used the former, given their higher sensitivity and resolution. These data are used to characterize the rest-frame optical morphology of the galaxies.

We extracted integrated photometry from the reduced \textit{HST} and \textit{JWST} images, and complemented it with additional mid- and far-IR data from Spitzer/IRAC and MIPS (\citealt{Dickinson2003,Whitaker:2014}) and Herschel/PACS (\citealt{Lutz:2011}), as well as (sub-)mm continuum measurements from our NOEMA$^{\rm 3D}$ observations (\citetalias{Chen2026}; see Table \ref{tab:photometry}). Using this multi-wavelength dataset, we performed integrated spectral energy distribution (SED) fitting with CIGALE \citep{Boquien:2019} to derive global galaxy properties, in particular stellar mass and star formation rate (SFR). The main assumptions adopted in the SED modeling are described in Appendix~A.

The resulting stellar masses and SFRs are listed in Table \ref{tab:ppts}. A more detailed analysis of the integrated and resolved SED properties of the full NOEMA$^{\rm 3D}$ sample will be presented in Tozzi et al. (in prep.).

\section{Kinematic modeling} \label{sec:analysis}

The main aim of this paper is to present the kinematic analyses of the galaxies studied in the \NOEMA survey. We present the kinematic maps and modeling for all galaxies. We also analyze and present the residual velocity maps (data - model), in a way similar to \citet{Genzel2023}. 
For nine galaxies, we interpret the residuals in terms of radial flows, and assess their spatial correlation with morphological structures (spiral arms and bars). 

Each galaxy is analyzed separately, in each of the available weightings (typically natural and robust). The residuals maps are only studied at the highest resolution, see Section \ref{sec:data}. Typically, the data obtained from the weightings that yielded the lowest spatial resolution and largest spatial extent are used to derive the kinematic PA, see more details in Section \ref{sec:PA}.

\subsection{Velocity map and masking}

We produced velocity and dispersion maps by fitting a gaussian to the CO line of each spectrum, on a pixel by pixel basis. We then computed the signal-to-noise ratio (S/N) for each pixel. To estimate the noise per channel, we calculate the standard deviation across the full cube, restricting the measurement to channels located more than 1 FWHM away from the inferred line center. We constructed an initial mask by excluding pixels with $\mathrm{S/N} < 3$. We subsequently apply a friends-of-friends algorithm, starting from the galaxy center, to remove isolated high-flux outliers located outside the main body of the galaxy.

We determine the center of each galaxy directly from \textit{JWST} (or \textit{HST}) imaging, typically using the reddest available filter where the galaxy centers are easily identifiable, by fitting a 2D Gaussian to the central light distribution. We then recenter all maps on these celestial coordinates and match them to a common spatial extent.  This allows for better accuracy than using the gas or dust maps, where the center might be less clearly identified. The adopted extent varies from source to source depending on the galaxy size and morphology.

\subsection{Derivation of position angle and inclination} \label{sec:PA}

The stellar mass maps are not always representative of the kinematic position angle (PA, or line of nodes). For example, the PA of G4\_38232, if derived solely from its near-IR light would be chosen at $\sim30^{\circ}$, while the kinematic PA is actually $\sim90^{\circ}$ offset. Lopsided galaxies, irregularities in the spiral arms, clumps or extinction will perturb the symmetry of the stellar maps and perturb the PA. Multiple methods are hence used to estimate the PA (positive north to east): 

\begin{itemize}
    \item We first measure the PA directly from the velocity map as the direction connecting the velocity extrema.
    \item In a similar way, we first identify the PA of the minor axis, and get the major axis from a $90^{\circ}$ offset.
    \item We fit the velocity and dispersion maps in 2D using \texttt{DysmalPy}, leaving the PA as a free parameter.
    \item We infer the PA from the isophotal fitting of both the NIR continuum light distribution (typically of the \textit{JWST} F444W filter, or the reddest available filter), and of the NOEMA CO flux data.
  \end{itemize}

\noindent
For methods relying on the NOEMA data (kinematic or integrated), we use preferentially the tapered cubes. Indeed these data provide the largest spatial extent and therefore yield more robust measurements. Their lower angular resolution also reduces sensitivity to small-scale perturbations and non-circular motions. While radial flows can in principle distort the velocity field and introduce local variations in the PA, the use of tapered (low-resolution) data minimizes these effects. In the end, we adopt a single PA for each galaxy, determined from the consistent convergence of all methods, providing a robust and well-constrained estimate.

The inclination is also a critical parameter as it is degenerate with most parameters derived through the modeling. To set the prior on inclination, we first perform isophotal fitting of the (typically) reddest \textit{JWST} filter available. In an iterative process we mask small clumps and asymmetric features, especially in the outer parts of galaxies, that may bias the fitting. The inclination is measured at the outer disk, where the isophotes are least affected by internal structure, and where the angle of the isophote major-axis matches the kinematic PA. Using the derived inclination as a first guess we fit the velocity map of the galaxy in 2D (with \texttt{DysmalPy}, with the lowest angular resolution CO data, to limit the effects of flows on the global kinematic patterns), leaving the inclination free. This further allows \texttt{DysmalPy} to fit the "spider-diagram" and gives us an extra validation of the expected inclination.

\subsection{Extraction of 1D kinematic profiles}

To extract the velocity and dispersion profiles used for modeling, we align a slit with the galaxy PA. We divide the slit into rectangular apertures whose width equals the synthesized beam major axis and whose height equals half that value. Adjacent apertures are spaced by half their height, resulting in partial overlap. We extract spectra by summing the flux within each aperture and fit each spectrum with a single Gaussian component. From these fits, we derive the velocity (line centroid), dispersion (linewidth, $\sigma$) and line flux (although not fitted) as a function of position along the slit. These measurements define the 1D profiles that we subsequently model and analyze (see Figure \ref{fig:vel_plots_all}).

\begin{figure*}[p]
    \renewcommand{\thefigure}{\arabic{figure}a}
    \centering
    \includegraphics[width=\linewidth,height=0.87\textheight,keepaspectratio]{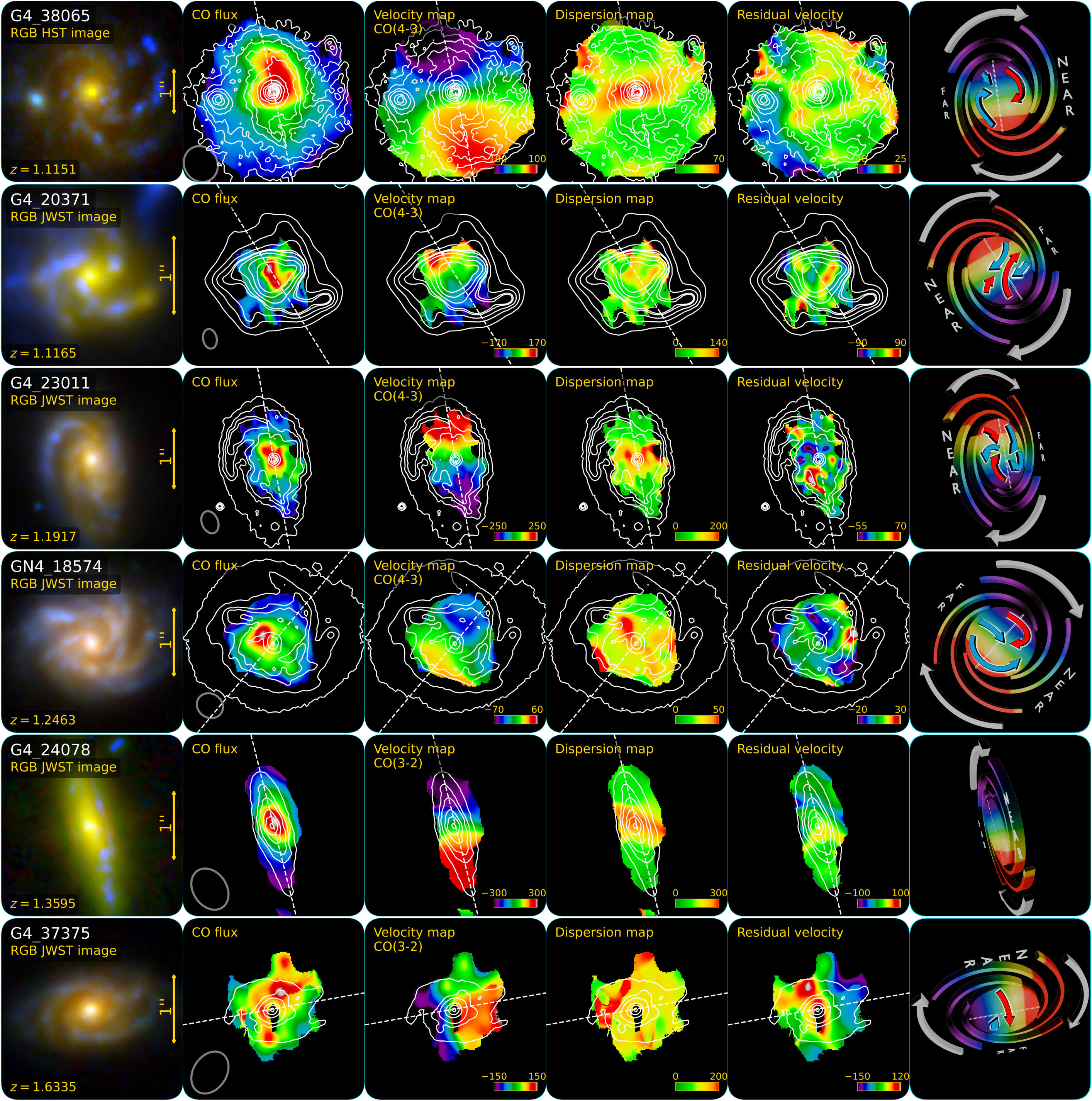}
    \caption{Overview of the different maps of the spiral galaxies in the sample. All maps --but the cartoon, last column-- share the same center and field of view. For each galaxy, from left to right: \textit{JWST} NIRCam color composite of the galaxy (\textit{HST} ACS \& WFC3 in the case of G4\_38065); CO flux map, integrated from the NOEMA data; Velocity map, with the NOEMA beam shape shown at the bottom left; Dispersion map; Velocity residuals (data - model); Cartoon representation of the galaxy with the inclination, PA and direction of the velocity field mimicking that of the galaxy. Gray arrows indicate the direction of rotation (assuming trailing spiral arms). Blue and red arrows within the galaxies' body  represent the observed coherent residual pattern, with the head of the arrow showing the dominant overall direction of the flow. 
    On each map the contours indicate a single \textit{JWST} or \textit{HST} filter (G4\_38065: 125W, G4\_37375: F200,G4\_17555: F200, G4\_20371: 444,G4\_23011: F150W,GN4\_18574: F200, GN4\_24517: F150W, G4\_24078: F277, G4\_38232: F200). The PA is shown as a white dashed line accross each map. A colorbar at the bottom of the kinematic maps shows the velocity extent and relation to color in \kms. Note that the orientation of G4\_24078 is unclear given the lack of a spiral structure to deduce the direction of rotation.}
    \label{fig:main_fig}
\end{figure*}

\begin{figure*}[!t]
\ContinuedFloat
    \renewcommand{\thefigure}{\arabic{figure}b}
    \centering
    \includegraphics[width=\linewidth]{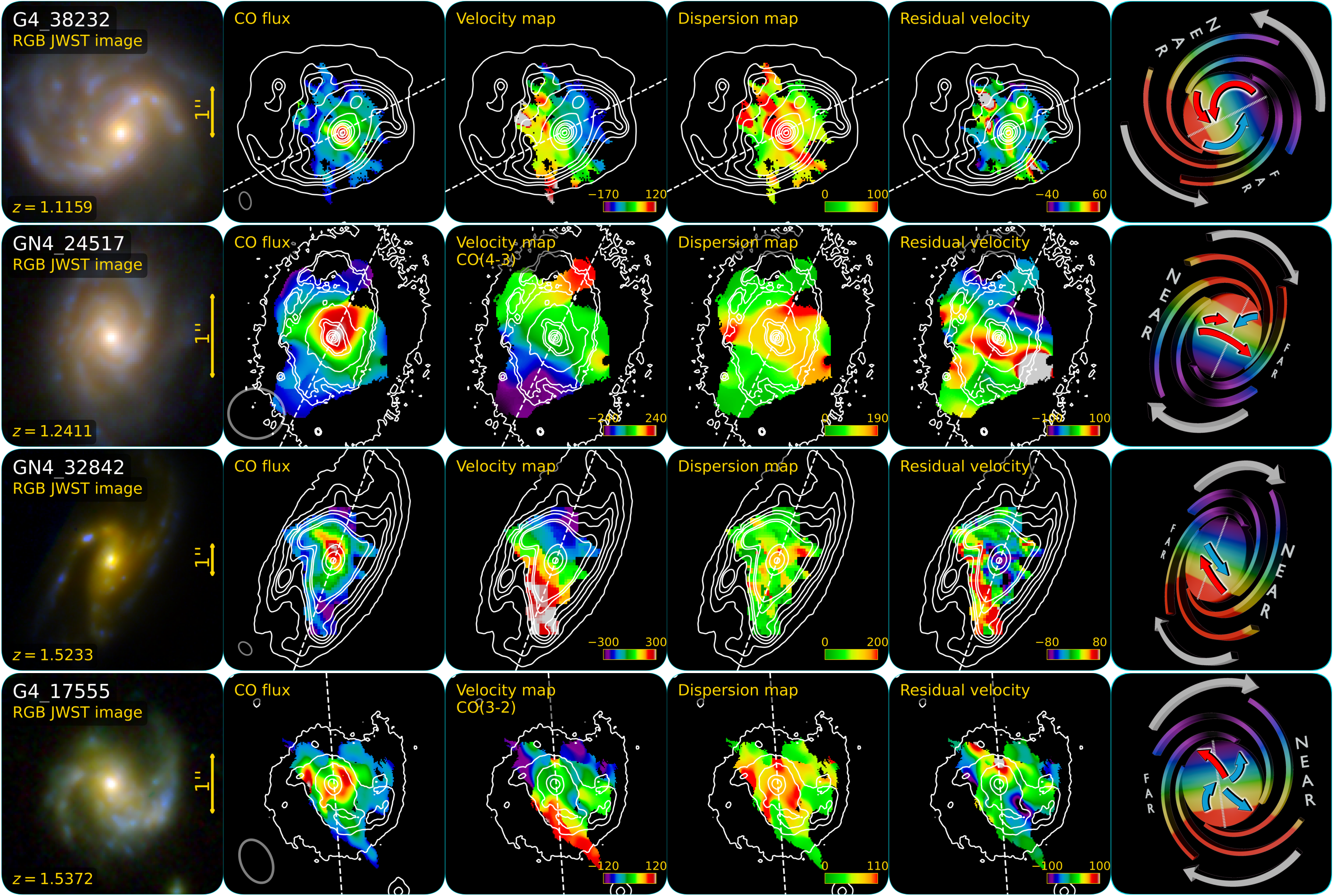}
    \caption{ Continued from Figure \ref{fig:main_fig}, for the remaining 4 galaxies, identified as barred. }
    \label{fig:2b}
\end{figure*}

\renewcommand{\thefigure}{\arabic{figure}}

\begin{table*}
\caption{\label{tab:kin_ppts}Parameters retrieved from the kinematic analysis with \texttt{DysmalPy}.}
\centering
\begin{tabular}{lccccccccc}
\hline\hline
Name & log(M$_{\rm bary}$/M$_\odot$) & $\sigma_0$ & $V_c(R_{\rm e})$& $V_c/\sigma_0$ & $f_{\mathrm{DM}}(R_{\rm e})$ & R$_\mathrm{e, disk}^\dagger$ & R$_\mathrm{e, bulge}^\dagger$ & Q$_{\mathrm{gas}}$\\
 &  & (\kms) & (\kms) & & &(kpc) & (kpc) &\\
\hline
G4\_38065 & $11.52^{+0.04}_{-0.04}$ & $18.3^{+0.75}_{-0.83}$ & 299 & $16.67^{+0.50}_{-0.42}$ & $0.18^{+0.11}_{-0.09}$ & 11.00$\pm$0.20 & 1.00$\pm$0.10 & 0.43 \\
G4\_38232 & $10.96^{+0.1}_{-0.1}$ & $20.0^{+5.84}_{-5.55}$ & 234 & $11.70^{+4.49}_{-2.64}$ & $0.04^{+0.22}_{-0.03}$ & 5.01$\pm$0.06 & 0.31$\pm$0.01 & 0.3 \\
G4\_20371 & $11.08^{+0.07}_{-0.09}$ & $49.8^{+3.29}_{-2.21}$ & 263 & $5.28^{+0.25}_{-0.33}$ & $0.09^{+0.11}_{-0.6}$ & 4.50$\pm$0.20 & 0.80$\pm$0.20 & 0.65 \\
G4\_23011 & $11.07^{+0.04}_{-0.06}$ & $35.1^{+3.3}_{-5.4}$ & $295$ & $8.42^{+1.53}_{-0.73}$ & $0.24^{+0.07}_{-0.09}$ & 4.00$\pm$0.05 & 0.35$\pm$0.04 & $0.60$ \\
GN4\_24517 & $10.93^{+0.14}_{-0.18}$ & $33.7^{+8.1}_{-16.2}$ & $333$ & $9.85^{+9.17}_{-1.90}$ & $0.54^{+0.14}_{-0.21}$ & 4.23$\pm$0.05 & 0.69$\pm$0.06 & $0.50$ \\
GN4\_18574 & $10.77^{+0.03}_{-0.04}$ & $25.8^{+0.7}_{-0.9}$ & $197$ & $7.65^{+0.28}_{-0.20}$ & $0.04^{+0.09}_{-0.02}$ & 3.70$\pm$0.10 & 0.30$\pm$0.20 & $0.39$ \\
G4\_24078 & $11.28^{+0.04}_{-0.06}$ & $38.7^{+10.0}_{-11.9}$ & $339$ & $8.77^{+3.91}_{-1.80}$ & $0.04^{+0.13}_{-0.02}$ & 3.08$\pm$0.02 & 0.30$\pm$0.03 & $0.46$ \\
GN4\_32842 & $11.30^{+0.16}_{-0.24}$ & $48.0^{+1.76}_{-2.78}$ & 379 & $7.90^{+0.49}_{-0.28}$ & $0.50^{+0.13}_{-0.13}$ & 6.24$\pm$0.09 & 0.60$\pm$0.08 & 0.78 \\
G4\_17555 & $10.68^{+0.16}_{-0.20}$ & $29.8^{+4.9}_{-6.4}$ & $249$ & $8.34^{+2.27}_{-1.18}$ & $0.55^{+0.20}_{-0.17}$ & 4.80$\pm$0.20 & 0.73$\pm$0.08 & $0.47$ \\
G4\_37375 & $10.80^{+0.10}_{-0.11}$ & $29.1^{+5.4}_{-4.5}$ & $248$ & $8.51^{+1.54}_{-1.33}$ & $0.34^{+0.17}_{-0.17}$ & 3.77$\pm$0.02 & 0.49$\pm$0.03 & $0.62$ \\

\hline
\end{tabular}
\tablefoot{
Columns from left to right: galaxy identifier; total baryonic mass from the kinematic modeling; isotropic velocity dispersion; circular velocity at one effective radius of the disk; ratio of circular velocity to velocity dispersion; dark matter fraction within an effective radius of the disk. Effective radius of the disk and effective radius of the bulge, obtained from photometric fitting, see \citetalias{Chen2026}. Toomre Q stability parameter of the gas, estimated following Eq. (3) of \citet{Genzel2014}, \citep[see also][]{Genzel2011, Genzel2023,Romeo2013,Uebler2019,Liu2023,Lee2025Cristal}. \\
$\dagger$: These parameters were fixed during the kinematic modeling.  
}
\end{table*}

\subsection{\texttt{DysmalPy} forward modeling}

We use the 1D profiles as inputs to generate galaxy models, using the publicly available  forward modeling code \texttt{DysmalPy}\footnote{\url{https://www.mpe.mpg.de/resources/IR/DYSMALPY/}} \citep{Davies2004a,Davies2004b,Cresci2009,Davies2011,Wuyts2016,Lang2017,Uebler2018,Price2021,Lee2025}. \texttt{DysmalPy} is a \texttt{Python} package designed to perform forward modeling of galaxy kinematics using multi-component mass models. It generates physically motivated rotating disk models and projects them into observable space, accounting for projection, beam smearing and spectral broadening. It has been widely tested to perform well in the ranges of S/N and angular resolution of our data (see references above), and includes a Markov Chain Monte Carlo (MCMC) minimization algorithm.
Each galaxy model that we generate with \texttt{DysmalPy} is axisymmetric and composed of a disk, a bulge and a NFW dark matter (DM) halo \citep{Navarro1996}. 

To limit the number of free parameters and reduce potential degeneracies we fix all quantities constrained through photometry. The effective radii (\Reff) of both disk and bulge, their Sérsic indices, as well as the bulge to total mass ratio (B/T), are fixed to values measured from photometric fitting or curve of growth fitting of the reddest filter (typically NIRCam F444W) and presented in \citetalias{Chen2026}. The DM concentration is fixed to $c = 10.9 \times (1+z)^{-0.83}$, following \citet{Genzel2020} and references therein. The intrinsic vertical to radial axis ratio of the disk is fixed to 1/5, appropriate for thick disks at $z\sim1$ \citep[see][]{Wuyts2016,Wisnioski2019,Lee2025Cristal}. The remaining parameters: total baryonic mass, DM fraction within the \Reff\, of the disk, isotropic velocity dispersion ($\sigma_0$), and systemic velocity, are left free. The total baryonic mass is bounded with 2 dex around the sum of the total stellar mass (inferred from SED fitting, see Section \ref{sec:SED}) and the gas mass derived in \citetalias{Chen2026}.

\texttt{DysmalPy} produces an axisymmetric mass model as a 4D hypercube, each cell of the hypercube containing the total model flux and full line-of-sight velocity distribution is then collapsed to a 3D data-cube convolved with the instrument PSF (here, the synthesized beam) and LSF ($\sigma\sim0.8\times$\,the channel size). A 1D profile is then extracted from the 3D data-cube and fitted to the data. From the best-fit 3D datacube we then extract the model velocity-map, in a similar fashion as done for the actual data. See the above references and \citet{Genzel2023} for more details on the functioning of \texttt{DysmalPy}. The kinematic properties of all galaxies as derived from our \texttt{DysmalPy} analysis are shown in Table \ref{tab:kin_ppts}.

It is worth noting that, as shown in \citet{Price2021} by comparing 1D and 2D analyses, most of the information about the rotation of the galaxy is contained within the 1D cut along the PA, and 1D analyses are less sensitive to the presence of non-circular motions. This is why we stick to the 1D method to derive the model and recover the kinematic properties of our galaxies. However, 2D analyses were also produced and yielded very consistent results.

As we generated cubes with different visibility weightings (see Section \ref{sec:data}), corresponding to different beam size and sensitivity to large-scale emission, we fit multiple 1D spectra simultaneously. Indeed, \texttt{DysmalPy} allows for simultaneous fitting of different observations (with different instrument characteristics) with a single galaxy model. The model is then convolved to each instrumental setup separately and the minimization is performed simultaneously on the different datasets. Figure \ref{fig:vel_plots_all} shows, for each galaxy in each studied weighting, the 1D data and corresponding model.

\subsection{Velocity residual maps and deprojection}

The final velocity residual maps are obtained by subtracting each beam-convolved model velocity map from the observed velocity maps. Both model and data are centered on the same coordinates, and the model is aligned to the galaxy PA. To interpret the residuals in terms of in-plane radial flows (inflowing or outflowing) we follow the procedure described in \citet{Pastras2025} (equation E33), considering pure radial flows. Essentially, residuals are interpreted as local inflows (negative in-plane radial velocities) or outflows (positive in-plane radial velocities) depending on (i) the direction of rotation of the galaxy (assuming trailing spiral arms) (ii) its inclination (iii) the side of the PA. The observed velocities residuals $u_{\text{resid}}$ are then scaled to true in-plane radial velocity depending on the angle $\phi$ from the major axis on the sky plane and the inclination $i$ of the galaxy. The in-plane radial velocities are hence defined as $u_r$ where:
\begin{equation} \label{eq:resid}
u_r=\frac{-u_{\text{resid}}}{sin\phi\tan i}\sqrt{\frac{\sin{}^2\phi}{\cos^2 i}+\cos^2\phi}    
\end{equation}

It should be noted that, for simplicity, we assume that the residuals stem purely from radial in-plane motions. However, in-plane tangential non circular motions can also contribute to the observed line-of-sight (LOS) velocity residuals, with such contributions being more significant in the cases of barred galaxies \citep[as discussed in][]{Pastras2025}. However this treatment allows for a rather straightforward interpretation of the results in terms of radial flows. To guide the reader's eye we produce 3D cartoons which mimic the galaxy's orientation, inclination, direction of rotation (clockwise or counter-clockwise), and direction of the velocity field. These cartoons subsequently show the side of the galaxy closer to the observer, blue and red arrows within the galaxies' body  representing the observed coherent residual pattern. The head of the arrows show the dominant overall direction of the in-plane gas flow, facilitating the interpretation of the observed patterns (see Figure \ref{fig:main_fig} last column).    

\subsection{Average in-plane radial velocities and mass flow rates} \label{sec:flow_rate_compute}

To compute the average inflow and outflow velocity for each galaxy (see Table \ref{tab:flows}), we consider only pixels with inferred in-plane radial velocities $|u_r|>10$\,\kms\,  
(see equation \ref{eq:resid}). This ensures the exclusion of pixels with flows with small deviation from pure axisymmetric rotation. we also remove all pixels too close to the major axis, where the LOS contribution of in-plane radial motions is minimal. Specifically, we mask all pixels within an angle $\phi_{min}$ of the PA, such that $u_r(\phi_{min})/u_{\mathrm{resid}}<0.1$. Furthermore, and to avoid including isolated pixels, we only include coherent groups of pixels of at least half the beam size in pixels. This leads to the selection of only coherent groups of inflowing (or outflowing) pixels. The average inflow and outflow are then computed separately, from all pixels in these groups, reported on Table \ref{tab:flows} and shown on their average absolute shown on Figure \ref{fig:inflow_outflow}. 

In addition to the average flow we compute the associated net mass flow rate following \citet{Pastras2025}: 
For each pixel we compute the corresponding molecular gas mass \citep[from the CO(3--2) or CO(4--3) flux, following][]{Tacconi2020}. This gas mass is then multiplied by the local in-plane radial velocity of each pixel. This map, in units of mass times velocity, is then integrated separately, in elliptical annuli with axis ratios derived from the inclination of the galaxy (where the widths of the annuli $w_{\mathrm{annulus}}= 1/4 \times b_{maj}$ (where $b_{maj}$ is the major axis of the beam). The integrated value inside each annulus is then divided by the annulus' width, and the total net flow rate is computed as the average over all annuli. As in \citet{Pastras2025}, the idea is that, in a perfectly steady state, a continuous flow toward the center would be at a constant rate throughout the galaxy. Results are presented on Figure \ref{fig:inflow_outflow}, lower panel.

{
\setlength{\tabcolsep}{4pt}

\begin{table}[ht] 
\caption{\label{tab:flows} Radial flows and net flow rates.}
\centering
\begin{tabular}{lccc}
\hline
Name  & $v_{r, in}$ & $v_{r, out}$ & $\dot{M}$ \\
& [\kms] & [\kms] & [\msun / yr]  \\

\hline
G4\_38065     & $-60.3 \pm 11.1$ & $+30.7 \pm 5.6$  &  $-54.0 \pm 15.5$ \\
G4\_38232     & $-121.2\pm18.5$ & $+103.6\pm22.2$  &  $-40.2\pm8$ \\
G4\_20371    &  $-72.1 \pm 12.5$  &  $+69.0 \pm 11.1$  &  $-152.0 \pm 30.4$ \\
G4\_23011   & $-70.7 \pm 10.1$ & $+44.7 \pm 5.7$  &  $-78.8 \pm 15.0$ \\
GN4\_24517   & $-121.7 \pm 17.9$ & $+114.3 \pm 15.4$  &  $-1.7 \pm 4.3$ \\
GN4\_18574    & $-74.7 \pm 31.3$ & $+42.5 \pm 17.6$  &  $-40.3 \pm 17.6$ \\
G4\_24078     & - & -  &  - \\
GN4\_32842   &  $-62.6\pm8.6$  &  $+60.1\pm10.3$   & $-128.0\pm25.6$  \\
G4\_17555    & $-97.1 \pm 19.7$ & $+130.8 \pm 27.7$ &  $32.0 \pm 8.3$ \\
G4\_37375     & $-107.5 \pm 14.2$ & $+140.2 \pm 18.5$ & $36.3 \pm 6.0$  \\
\hline
\end{tabular}
\tablefoot{Radial inflow/outflow in the galaxy plane, and net flow rate. $v_{r, in}$ is the average radial inflow velocity in the plane of the galaxy. $v_{r, out}$ is the average radial outflow velocity in the galaxy plane. $\dot{M}$ is the net flow rate, see Section \ref{sec:flow_rate_compute} for a detailed description. Negative (positive) net flow rates correspond to inflowing (outflowing) mass rates. }
\end{table}
}

\section{Results} \label{sec:results}

\subsection{Kinematic properties: \NOEMA galaxies as turbulent rotating disks}

\begin{sidewaysfigure*}[ht]
    \centering
    \includegraphics[width=0.9\linewidth]{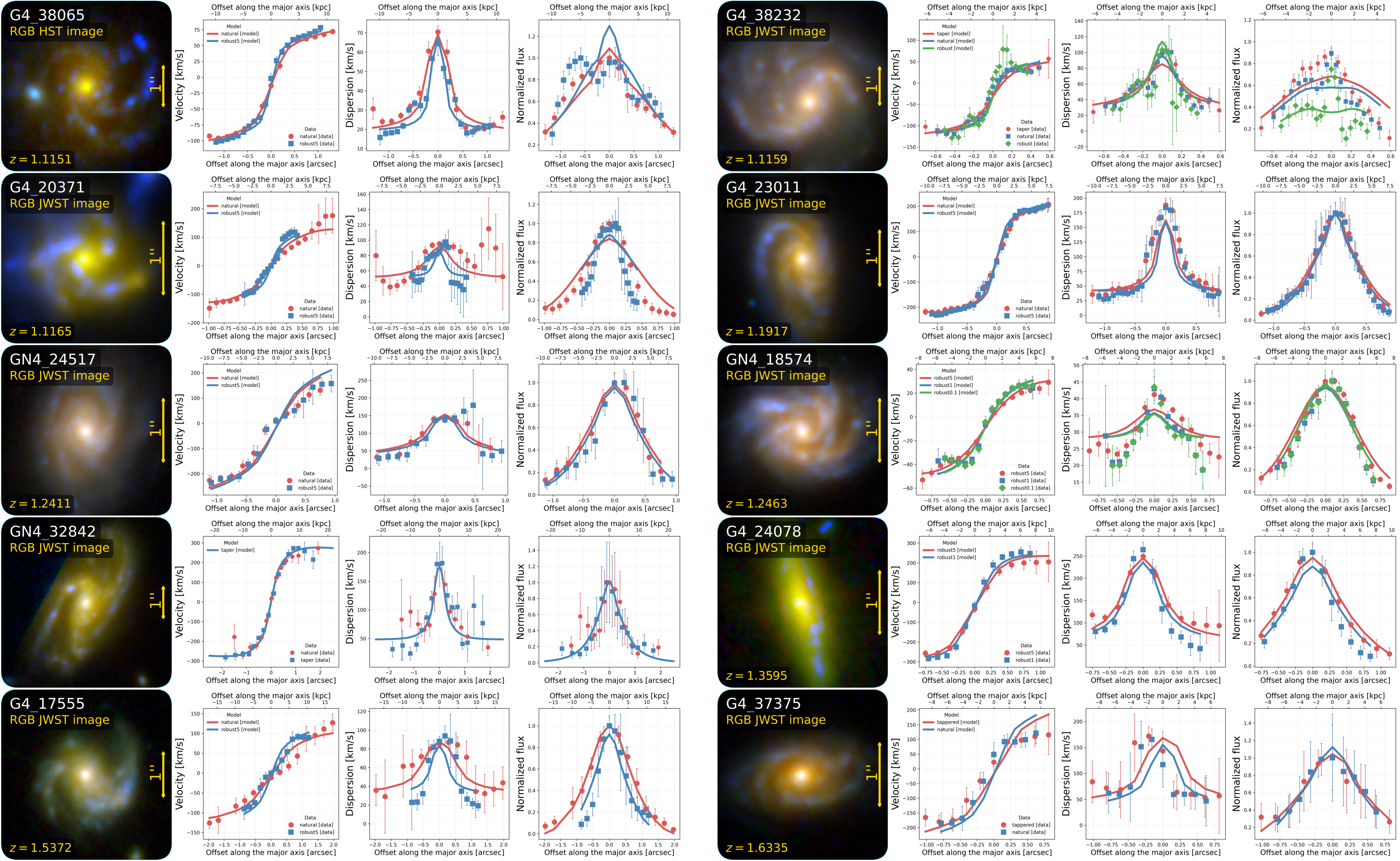}
    \caption{Kinematic profiles of the \NOEMA galaxies (one per row). From left to right: color composite image of the galaxy; rotation curve, data and model; dispersion profile, data and model; CO flux profile, data and model. All profiles are extracted from a slit along the major axis (see Section \ref{sec:analysis}). The different weightings of each galaxy are fitted  simultaneously to increase reliability.}
    \label{fig:vel_plots_all}
\end{sidewaysfigure*}

All galaxies studied in this analysis present regular rotation curves (RC; see Figure \ref{fig:vel_plots_all}) and can be well characterized up to a few $R_e$ (2.74 on average, from weightings with the largest extent). Most RCs appear to flatten at an average turnover radius $\overline{R_t}\sim 1.04\, R_e$, while a few continue to rise. From the \texttt{DysmalPy} analysis we infer an average dark matter fraction within $R_e$ of $\overline{\mathrm{f_{DM}}(R_e)}\sim0.26\pm0.20$. This is in good agreement with typical values expected at this redshift and in this mass range (e.g. \citet{Genzel2020,NestorShachar2023}, see also \citealt{Danhaive2026}).

\begin{figure}
    \centering
    \includegraphics[width=1\linewidth]{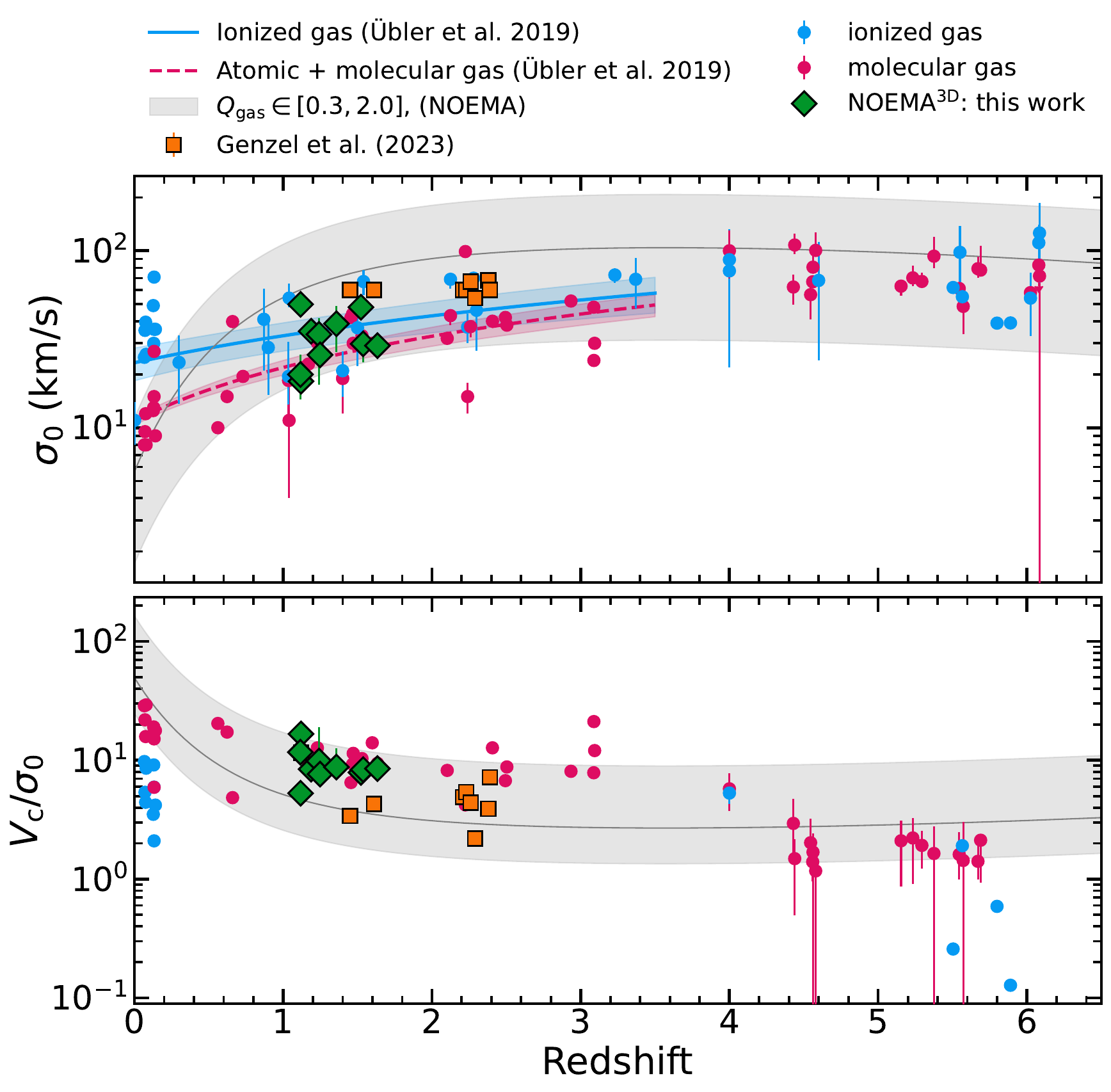}
    \caption{(top) Velocity dispersion as a function of redshift. (bottom) Rotation velocity normalized by the velocity dispersion at \Reff\, as a function of redshift. The gray shaded area encloses the corresponding range for the Toomre parameter Q $\in$ [0.3, 2.0], at the \NOEMA sample average of log(M$_\star$/M$_\odot)=10.9$ and $V_c=280$\,\kms. Data points from the \NOEMA galaxies are shown in green, galaxies from \citet{Genzel2023} are shown in orange. Additional data points are extracted from the literature: \citet{Hodge2012,Turner2017,Johnson2018,Girard2019,Girard2021,Uebler2019,Uebler2024c,Price2021,Lelli2023,Liu2023,Liu2024,Parlanti2023,Parlanti2024,Rizzo2023,Mai2024,Danhaive2025,deGraaf2025,Lee2025Cristal}. Relations derived in \citet{Uebler2019} for the ionized and atomic + molecular gas are shown in blue and pink, respectively.}
    \label{fig:sigma_evo}
\end{figure}

From the analysis of the velocity dispersion profiles, we observe that the \NOEMA galaxies are turbulent (with a median $\sigma_0^{med}\sim 32\pm10$\kms, ranging from $\sigma_0\sim18.3$\kms\, to $\sigma_0\sim49.8$\kms, after correction for beam smearing and observational effects), as expected for star-forming galaxies at this redshift (see Figure \ref{fig:sigma_evo}). This confirms the overall turbulent nature of molecular gas disks at $z\sim1.3$, contrasting with the samples at $z\sim2$, with even higher average dispersions, as expected from their higher gas fractions and gas accretion rates. The velocity dispersion profiles are typically regular, with a strong central peak and a rapid transition to a constant profile (see Figure \ref{fig:vel_plots_all}). For some galaxies, the dispersion maps seem to reach slightly higher values on average in regions showing signatures of velocity residuals (see Figures \ref{fig:main_fig} and \ref{fig:2b}). This can be expected due to the potentially enhanced turbulence in regions of flows, or if the gas splits into multiple components, e.g. some gas flowing inward while some maintains ordered rotation. 

When compared to their rotation velocity, the \NOEMA galaxies, while turbulent, are rotation dominated with a median $(V_c/\sigma_0)^{\mathrm{med}}\sim8.64\pm2.89$. Overall, the \NOEMA galaxies have $\sigma_0$ values comparable to those observed at this redshift, where the molecular gas dispersion is typically lower than that of ionized gas at the same lookback time. For example, \citet{Uebler2019} find velocity dispersion on average of 32.5--38.2 \kms, for ionized gas in the redshift range $z\sim1-1.5$. This difference could reflect the distinct phases traced by molecular and ionized gas, with the latter probing more diffuse and dynamically hotter components. However it could also be due to differences in the mass selection. Compared to the sample studied in \citet{Genzel2023}, at slightly higher redshifts, the measured dispersions are consistent with an increase with lookback time, as predicted in \citet{Uebler2019}. The observed dispersions are also consistent with marginally stable disks ($Q_{\mathrm{gas}}\sim1$), although typically slightly lower, supporting a picture in which turbulence is regulated by gravitational instability. Q values close to unity are consistent with the detection of spiral structures in our galaxies. Spiral arms in turn can drive the radial flows observed. The presence of significant turbulence may also facilitate radial gas transport, as it can enhance angular momentum redistribution and contribute to the observed non-circular motions. Our observations however indicate that arms and bars most likely dominate.

We note that the dispersion peak is sometimes slightly underestimated by our models, which might be due to the presence of a central SMBH whose impact could be underestimated by our disk+bulge modeling approach. This effect is, however, mild and does not significantly affect our analysis.

\subsection{High residual velocities as evidence of strong radial flows}

Analyses of the ten galaxies presented in this study are grouped into two main categories: 
(i) \textbf{clear residuals along the spiral arms}, and 
(ii) \textbf{an apparent streaming pattern in barred galaxies}.

\begin{figure}
    \centering
    \includegraphics[width=0.95\linewidth]{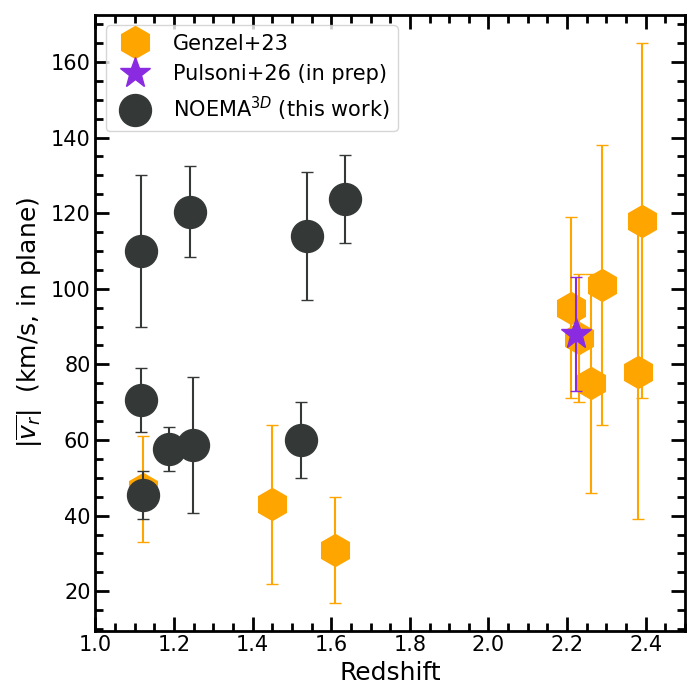}
    \includegraphics[width=0.95\linewidth]{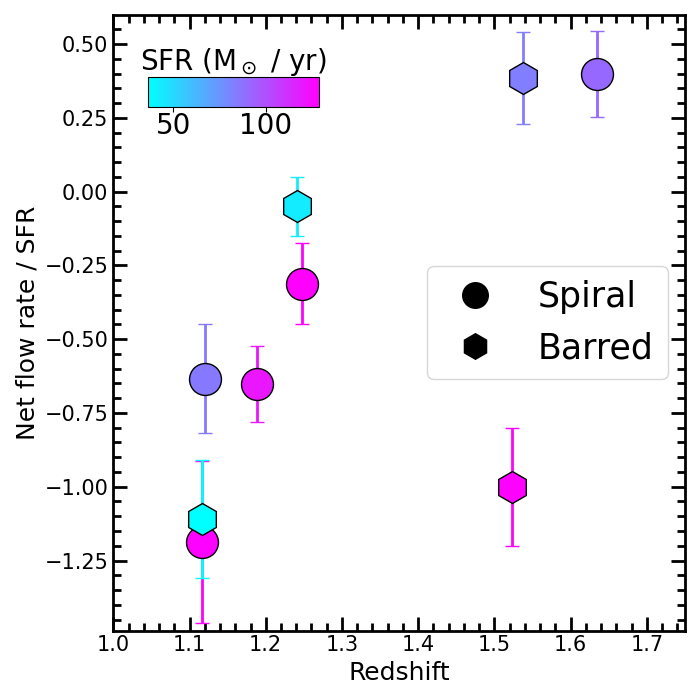}
    \caption{(top) Average in plane radial flow velocities as a function of redshift, assuming residual velocities represent purely radial flows (see Section \ref{sec:flow_rate_compute}). 
    Points in orange show the velocities measured in \citet[][]{Genzel2023}. (bottom) Net flow rates normalized by the SFR of each galaxy, as a function of redshift. Spiral galaxies are indicated with a circle while barred galaxies are indicated with a hexagon. Data points are color coded for the galaxy SFR. Errors are computed by propagating errors from the velocity map and error on inclination (top) as well as error on the flux (bottom).}
    \label{fig:inflow_outflow}
\end{figure}

\begin{itemize}

\item[(i)] 
For four of the galaxies (G4\_38065, G4\_20371, G4\_23011 and GN4\_18574) we identify clear velocity residuals in regions associated with the spiral arms (or inter-arm regions). \textbf{Assuming pure radial flows we observe a mix of inflows and outflows, although with a clear predominance of inflows with negative net flow rates} (see Figure \ref{fig:inflow_outflow}). 
\textbf{The typical velocities of the residuals are high, of order 50--100 \kms\,in the plane of the galaxies}, similar to the ones observed in \citet{Genzel2023} and in Pulsoni et al., in prep. The flow velocities are much higher than in local galaxies \citep[e.g.][]{DiTeodoro_2021}. This is likely a direct consequence of the higher gas fractions observed at cosmic noon and beyond, and highlights the tight link between the high gas fractions and the fast growth of the galaxies. In addition, \textbf{we observe high net flow rates, on the order of the SFR ($\overline{\dot{M}}\sim-50$ \msun/yr}), demonstrating how the \textbf{spiral arms can play an instrumental role in accreting the gas necessary for star formation, and funneling it inwards, toward the bulge and SMBH.}

\item[(ii)] 
In the full sample, we identify four galaxies as barred: GN4\_32842, G4\_38232, G4\_17555, and GN4\_24517. Detailed studies of the first two are presented in separate papers (\citet{Pastras2025}, and in prep.). In all four, an apparent “inflow-outflow” pattern is identified in the residuals, which can be interpreted as gas flows in the presence of a bar and possibly bar lanes / dust lane shocks \citep[e.g.][]{Pastras2025}. The exact interpretation of these residual patterns requires careful analysis and a combination of radial and tangential flows. The simpler approach followed in this paper will not capture these more complicated residual patterns. We can still, however, extrapolate from the results of the analysis of GN4\_32842 to estimate the order of magnitude of the flows around the bars of the other galaxies in the sample. Indeed, both G4\_17555 and GN4\_24517 \textbf{show residual patterns  resembling the ones seen in GN4\_32842 and G4\_38232, with comparable average residual velocities. It is thus likely that the in-plane radial velocities are large, of the order of $\sim50-100$\,km s$^{-1}$, and result in inflow rates of the order of the galaxy-integrated SFR \citep{Pastras2025}.} This is much larger than the expected bar-mediated azimuthally-averaged inflow rates of $<5$ \kms\,from simulations of local-like barred galaxies \citep[e.g.][]{Athanassoula1992}.

\end{itemize}

These findings underscore the critical role of internal dynamical processes, specifically spiral arm and bar-driven gas flows, in the evolution of galaxies and their mass assembly. The detection of the high-velocity residuals (50–100 \kms), driving inflow rates of the order of the star formation rate, reveals that spiral arms and bars are not passive features, but play a crucial role in the channeling of gas from the outer disk toward the central region. This efficient fueling mechanism directly supports the sustained growth of stellar bulges and supermassive black holes, highlighting the intimate connection between morphology and galactic growth. These results emphasize that at $z\sim1.5$, galaxies are dynamically active systems, where internal gravitational instabilities in spiral arms and bars are key drivers of gas accretion, fueling star formation, and enabling the rapid growth of both stellar and black hole masses. This reinforces the need to incorporate the impact of non-axisymmetric structures on gas kinematics to fully understand how galaxies assembled their mass around the peak of cosmic star formation.

\section{Discussion} \label{sec:discussion}

While it is well established \citep[e.g.][]{Roberts1979,Athanassoula1992,Bournaud2002,Combes2008,Pastras2025} that non-circular motions at the end of and along galactic bars will typically be the sum of radial motions and deviations from axisymmetric rotation in the azimuthal direction (tangential streaming motions), the analysis performed here focused solely on radial motions. In the non-barred galaxies analyzed here, one can see that the relative contribution of the velocity residuals along the minor-axis (where only radial flows contribute) to the total residuals, compared to that of the residuals along the major-axis (where only tangential flows will contribute) always favor the former. This was also observed in \citet{Genzel2023}, where flows are preferably observed along (or close to) the minor axis. For this reason, and to facilitate the computation of global flow rates, we follow the typical assumption that radial motions will dominate. However, this assumption should be regarded as an approximation. 

In general, when gas orbits deviate from circularity, the decomposition of the observed line-of-sight velocity into radial and tangential components becomes non-trivial. Away from the principal axes, both components contribute simultaneously, and their relative weights depend on the (generally unknown) geometry of the underlying streamlines. In particular, without explicit knowledge of the orientation and ellipticity of the gas orbits, it is not possible to uniquely separate the projection of tangential streaming motions from that of radial flows. As a result, part of the signal attributed here to radial motions may in fact arise from tangential components projected along the line of sight. This limitation is mitigated by focusing on regions away from the major axis, where the contribution from tangential motions is lower. Indeed, shocks, if present, will have a dominant contribution close to the major axis, which is why a cone around the major axis is excluded when computing the flow rate (see Pastras et al., in prep). Consequently, as measurements are confined to regions away from the major axis, a fraction of the total radial flow may also remain unaccounted for. The net radial flow rates derived in our analysis could hence be overestimated due to the neglect of the contribution of tangential streaming motions. This is especially relevant in the limiting case of highly elongated but closed streamlines, where non-zero radial velocities may be measured locally despite the absence of a net mass transport. \citet{Salibur2026} show for example that projection effects of elliptical orbits can be mistaken for radial flows along bars. Alternative approaches, such as torque-based methods, could in principle provide a more complete characterization of gas flows, as demonstrated in Pastras et al., in prep for one of the \NOEMA targets.

In local spiral galaxies, and according to density wave theory, inflows are observed along the arms while outflows are observed in the inter-arm region \citep[e.g.][]{Maciejewski2004,Davies2009}. In our work we note that the spatial resolution attained by NOEMA is not sufficient to distinguish between these two patterns, and the overall velocity we do observe could be the sum of both flows, going in opposite directions. If so, this would directly imply a underestimation of the derived radial velocities, which could hence be even faster than the ones observed. 
Furthermore, one should note that the exact location of the velocity residuals observed here --whether on the arm or in the inter-arm-- may be uncertain due to limited spatial resolution.

Overall the analysis highlights the clear presence of high-velocity residuals in all the galaxies. For some galaxies we see a clear mix of residuals interpreted as inflows and outflows, for others inflows seem to dominate. While some cases might be harder to interpret, others showcase very clear residual patterns, that can be clearly linked to the arms or bar. G4\_38065, for example, exhibits very strong and fast residuals on its western side, following very clearly the spiral pattern. It also seems to be associated with a bright clump, which might be infalling. The variety of signatures observed showcases that the relative strength and prevalence of the phenomena described in this analysis may vary from galaxy to galaxy. They could depend on their morphology, environment, or physical characteristics (e.g. mass, size etc.). However, it is evident from the sample described here that the objects with the better S/N, and the better ordered kinematic fields, are also the ones where the residuals, and their correspondence with morphological features, are strongest. These results suggest that fast radial flows along spiral arms or bars may be a common feature of galaxies at cosmic noon, although their detection requires deep integrations with high S/N and high spatial resolution.  

Warps are relatively common in the outer disk of local galaxies \citep[through HI observations, also of the Milky-Way, e.g.][]{Levine2006,vanderKruit2011}, and can cause significant deviations from pure circular motions that, furthermore, could follow the spiral arms. Nevertheless, the large velocity dispersions observed at cosmic-noon (and in the \NOEMA galaxies), and their expected isotropy \citep[e.g.][]{Genzel2011,vanDokkum2015,Wisnioski2015,Uebler2019}, should make disks stable against buckling type instabilities \citep{Toomre1964,Merritt-Sellwood1994}, see section 4.2, of \citet{Genzel2023} for a more detailed discussion on warps and tidal effects. Furthermore, and as observed in \citet{Tsukui2024}, the signature of such bending of the disk would be a rise of the rotation curve on one side, and a drop on the other side, which we do not observe in the \NOEMA galaxies. 

Looking ahead, if fast radial flows along spiral arms or bars are indeed a common feature of galaxies at these epochs and beyond, their systematic identifications will require a significant investment of telescope time to reach the required signal-to-noise ratio and spatial resolution. In particular, probing the detailed dynamics and assessing the role of gravitational torques systematically, demands observations capable of resolving sub-kiloparsec structures with high fidelity and with a high spectral resolution. This is demanding for current existing facilities, requiring dedicated studies with long integrations on single objects, like the one presented here. Such capabilities will be essential for confirming that these radial flows are ubiquitous and for integrating them into a coherent framework of galaxy evolution at cosmic noon and beyond.

\section{Conclusion / Summary} \label{sec:ccl}

In this paper we presented \NOEMA, a unique study focusing on the molecular gas kinematics of ten massive main sequence galaxies just after Cosmic Noon ($z\sim1.1-1.6$). We described the data acquisition, reduction and the sample properties. We obtained high resolution (typical resolution of $\sim0.47''$) deep CO observations (typical rms $\sim0.13$\,mJy/beam), from which we derived detailed kinematic properties of the galaxies. Using the kinematic forward modeling tool \texttt{DysmalPy} we constructed robust kinematic models, from which we then derived the residual velocity maps, revealing the presence of substantial planar non-circular motions in the galaxies. Using the available multi-band \textit{JWST} imaging of the restframe-NIR continuum emission (probing the bulk of the stellar component in our targets’ disks), we interpret the identified motions, as inflows driven by the non-axisymemtric structure, i.e. spirals and bars. Overall, our results can be summarized as follows:

\textbf{From the CO data we derive high-quality kinematic maps} in multiple weightings, from the most extended and lowest resolution tapered data to the highest resolution, more compact, robust-weighted data. \textbf{All galaxies show clear signs of ordered rotation}, as expected from their observed morphologies, \textbf{with relatively high level of turbulence} ($\sigma_0^{\mathrm{med}}\sim32\pm10$\,\kms, $(V_c/\sigma_0)^{\mathrm{med}}\sim8.64\pm2.89$). 

After generating a kinematic model with \texttt{DysmalPy}, we subtract the model velocity maps from the measured ones and find clear velocity residuals. \textbf{The residuals are spatially coherent, forming large consistent patches. They are fast, revealing the presence of in-plane velocities of order $\sim50-100$\kms\.}   

\textbf{We interpret these flows as being mostly radial, as they are typically maximal along the minor axis.} Using the rotation direction of the galaxy (considering trailing spiral arms) we infer the side of the galaxy closest to the observer and hence deduce the direction of the radial flow: inflowing or outflowing. We then derive at each point of the galaxy the true in-plane radial velocity from the observed residuals. 

We typically see a mix of inflows and outflows, however the \textbf{inflows seem to dominate overall in unbarred spiral galaxies.} More specifically, the cases with the best S/N are clearly inflow dominated. \textbf{In barred galaxies, an apparent “inflow-outflow” pattern can be identified,} in overall agreement with expectations for bar-driven gas flows, as indicated by separate detailed studies of some of our barred targets (\citet{Pastras2025}, and in prep.).

\textbf{Most importantly the residuals typically align well with the presence of the non-axisymmetric structures (spiral arms and bars), showcasing their strong intrinsic relation.} These observations are only possible thanks to the sensitivity and very high-resolution of the NOEMA kinematic observations, combined with deep, multi-band \textit{JWST} imaging.

Using the CO line flux maps, combined with the in-plane radial velocity maps, we compute the average flow rate of the galaxies. \textbf{The flow rates are typically negative (inflow dominated), with an average flow rate  $\overline{\dot{M}}\sim-50$ \msun/yr, of order of the SFR}, although with a large spread. \textbf{The result underscores the efficiency with which spiral arms and bars channel molecular gas toward the central regions, rapidly transporting star-forming fuel from the outer disk to the central regions.} 

Our observations and analysis allowed us to put into light the presence of strong non-circular motions following non-axisymmetric morphological structures in star-forming galaxies at cosmic noon, and their possible role as drivers of the evolution of their host galaxies. We showed that if they are interpreted as radial flows, they are predominantly inflows, reach very high velocities (much faster than in local galaxies), and are associated with large amounts of gas transport. These flows would be sufficient to fuel the high SFR of galaxies at cosmic noon, promoting bulge formation and possibly the feeding of central SMBHs.

\begin{acknowledgements}
We thank the IRAM staff and the IRAM Partnership (MPG in Germany, INSU/CNRS in France and IGN in Spain) for their hard work and support of the NOEMA upgrade at IRAM, which made the \NOEMA project possible. 
CB, GT, JC, JMES, LL and NMFS  acknowledge funding by the European Union (ERC Advanced Grant GALPHYS, 101055023).  
GM and HÜ acknowledge funding by the European Union (ERC APEX, 101164796).
Views and opinions expressed are, however, those of the author(s) only and do not necessarily reflect those of the European Union or the European Research Council. Neither the European Union nor the granting authority can be held responsible for them.
HÜ thanks the Max Planck Society for support through the Lise Meitner Excellence Program.
TN acknowledges the support of the Deutsche Forschungsgemeinschaft (DFG, German Research Foundation) under Germany’s Excellence Strategy - EXC-2094 - 390783311 of the DFG Cluster of Excellence ‘ORIGINS’. 
SGB and AU acknowledge support from the Spanish grant PID2022-138560NB-I00, funded by MCIN/AEI/10.13039/501100011033/FEDER, EU. 
Some of the data products presented herein were retrieved from the Dawn \textit{JWST} Archive (DJA). DJA is an initiative of the Cosmic Dawn Center (DAWN), which is funded by the Danish National Research Foundation under grant DNRF140.

\end{acknowledgements}


\bibliography{myBib}

\begin{appendix}

\section{Photometry}

The SFR and stellar masses shown in Table \ref{tab:ppts} are derived from integrated SED fitting of all photometry available for each galaxy (see Table \ref{tab:photometry}). More detailed SED analysis, including resolved SED fitting, will be presented  in a separate paper (Tozzi et al., in prep.). 

For the integrated SED fitting, we used \texttt{CIGALE} \citep{Boquien:2019}, fixing the redshift to the spectroscopic value for each galaxy. We assumed an exponentially declining star formation history with an e-folding time $\tau = 8000$ Myr, effectively corresponding to a nearly constant SFH over the timescales of interest. We adopted \citet{Bruzual:2003} stellar population synthesis models with a \citet{Chabrier:2003} initial mass function and fixed solar metallicity ($Z = 0.02$), including nebular emission. Dust attenuation was modeled using the \citet{Calzetti:2000} law, while dust emission was reproduced using the templates of \citet{Dale:2014}. 

Star formation rates were derived from the combination of UV and IR luminosities of the best-fit SEDs following \citet{Kennicutt:1998}. Stellar masses correspond to the direct output of \texttt{CIGALE}.

\begin{table*}[]
\centering
\caption{Available photometry for each galaxy in the sample used for the integrated SED fitting, comprising data from \textit{HST}/ACS and WFC3/IR, \textit{JWST}/NIRCam and MIRI, \textit{Spitzer}/IRAC and MIPS, \textit{Herschel}/PACS, and NOEMA (band 2 or band 3). For the spatially resolved SED fitting, only \textit{HST}/ACS and WFC3/IR, and \textit{JWST}/NIRCam imaging were used.}

\label{tab:photometry}
\setlength{\tabcolsep}{2pt}
\renewcommand{\arraystretch}{1.2}

\begin{adjustbox}{center,raise=0pt}
\begin{tabular}{cccccccccccc}
\hline
Instrument & Band & G4\_17555 & G4\_20371 & G4\_23011 & G4\_24078 & G4\_37375 & G4\_38065 & G4\_38232 & GN4\_18574 & GN4\_24517 & GN4\_32842 \\
\hline
\multirow{4}{*}{\shortstack{\textbf{HST} \\ ACS}} 
 & F435W  & \tick & \tick & \tick & \tick & \tick & \tick & \tick & \tick & \tick & \tick \\
 & F606W  & \tick & \tick & \tick & \tick & \tick & \tick & \tick & \tick & \tick & \tick \\
 & F775W  &  &  &  &  &  &  &  & \tick & \tick & \tick \\
 & F814W  & \tick & \tick & \tick & \tick & \tick & \tick & \tick & \tick & \tick & \tick \\
\hline
\multirow{4}{*}{\shortstack{\textbf{HST} \\ WFC3/IR}} 
 & F105W  &  &  &  &  &  &  &  & \tick & \tick & \tick \\
 & F125W  & \tick & \tick & \tick & \tick & \tick & \tick & \tick & \tick & \tick & \tick \\
 & F140W  & \tick &  & \tick & \tick &  & \tick & \tick & \tick & \tick & \tick \\
 & F160W  & \tick & \tick & \tick & \tick & \tick & \tick & \tick & \tick & \tick & \tick \\
\hline
\multirow{11}{*}{\shortstack{\textbf{JWST} \\ NIRCam}} 
 & F090W  & \tick & \tick & \tick & \tick & \tick &  & \tick & \tick & \tick & \tick \\
 & F115W  & \tick &  & \tick & \tick & \tick &  & \tick & \tick & \tick & \tick \\
 & F150W  & \tick &  & \tick & \tick & \tick &  & \tick & \tick & \tick & \tick \\
 & F182M  &  &  &  &  &  &  &  & \tick & \tick &  \\
 & F200W  & \tick &  & \tick & \tick & \tick &  & \tick & \tick & \tick & \tick \\
 & F210M  &  &  &  &  &  &  &  & \tick & \tick &  \\
 & F277W  & \tick &  & \tick & \tick & \tick &  & \tick & \tick & \tick & \tick \\
 & F335M  &  &  &  &  &  &  &  & \tick & \tick & \tick \\
 & F356W  & \tick &  & \tick & \tick & \tick &  & \tick & \tick & \tick & \tick \\
 & F410M  & \tick &  & \tick & \tick & \tick &  & \tick & \tick & \tick & \tick \\
 & F444W  & \tick & \tick & \tick & \tick & \tick &  & \tick & \tick & \tick & \tick \\
\hline
\multirow{7}{*}{\shortstack{\textbf{JWST} \\ MIRI}} 
 & F560W  &  &  &  &  &  &  &  &  & \tick &  \\
 & F770W  & \tick &  & \tick &  & \tick &  & \tick &  & \tick &  \\
 & F1000W & \tick &  & \tick & \tick & \tick &  & \tick &  & \tick &  \\
 & F1280W & \tick &  &  & \tick &  &  &  &  &  &  \\
 & F1500W & \tick &  & \tick & \tick & \tick &  & \tick &  &  &  \\
 & F1800W &  &  &  & \tick &  &  &  &  &  &  \\
 & F2100W & \tick &  & \tick &  & \tick &  & \tick &  & \tick &  \\
 \hline
\multirow{4}{*}{\shortstack{\textbf{\textit{Spitzer}} \\ IRAC}} 
 & 3.6~$\rm \mu m$  &  & \tick &  &  &  & \tick &  &  &  &  \\
 & 4.5~$\rm \mu m$  &  &  &  &  &  & \tick &  &  &  &  \\
 & 5.7~$\rm \mu m$ & \tick & \tick & \tick & \tick & \tick & \tick & \tick & \tick &  & \tick \\
 & 8.0~$\rm \mu m$ &  & \tick &  & \tick &  & \tick & & \tick &  & \tick \\
\hline
\multirow{1}{*}{\shortstack{ MIPS}} 
 & 24~$\rm \mu m$  & \tick & \tick & \tick & \tick & \tick & \tick & \tick & \tick & \tick & \tick \\
 \hline
 \multirow{2}{*}{\shortstack{\textbf{\textit{Herchel}} \\ PACS}} 
 & 100~$\rm \mu m$  &  &  & \tick & \tick & \tick & \tick &  & \tick & \tick & \tick \\
 & 160~$\rm \mu m$ &  & \tick & \tick & \tick &  & \tick &  & \tick &  & \tick \\
 \hline
\multirow{2}{*}{\shortstack{\textbf{NOEMA}}} 
& 1.3~mm  &  & \tick & \tick &  &  & \tick & \tick & \tick & \tick &  \\
 & 2~mm  & \tick &  &  & \tick & \tick &  &  &  &  & \tick \\
\hline
\end{tabular}
\end{adjustbox}

\end{table*}

\end{appendix}

\end{document}